\documentclass[aip,adv,graphicx,reprint,floatfix,showkeys,a4paper]{revtex4-1}

\usepackage{inputenc}
\inputencoding{utf8}
\usepackage[OT1]{fontenc}
\usepackage{siunitx}
\usepackage{graphicx}
\usepackage{listings}
\usepackage{amssymb,amsmath}
\usepackage{array}
\usepackage{hyperref}
\usepackage{color}
\usepackage{textcomp}

\newcommand*{\diff}[2]{\frac{\mathrm{d} #1}{\mathrm{d} #2}}
\newcommand*{\TdTdt}{$\left(T,\frac{\mathrm{d} T}{\mathrm{d}t}\right)$}
\newcommand*{\vdvdt}{$\left(v,\frac{\mathrm{d} v}{\mathrm{d} t}\right)$}
\newcommand*{\Tg}{$T_\text{g}$}
\newcommand*{\TgTC}{$T_{\text{g},\text{TC}}$}
\newcommand*{\Tm}{$T_\text{m}$}

\begin{document}

% number on the title page in preprint mode.  Multiple \preprint
% commands are allowed.  \preprint{}
 
\title{Thermalization Calorimetry: A simple method for investigating
  glass transition and crystallization of supercooled liquids}
\author{Bo Jakobsen} \email{boj@dirac.ruc.dk} \author{Alejandro Sanz}
\author{Kristine Niss} \author{Tina Hecksher}  \author{Ib
  H. Pedersen} \author{Torben Rasmussen} \author{Tage Christensen}
\author{Niels Boye Olsen} \author{Jeppe C. Dyre} \affiliation{``Glass and Time'',
  IMFUFA, Department of Sciences, Roskilde University, Postbox 260,
  DK-4000 Roskilde, Denmark}

%\date{\today}

\begin{abstract}
  We present a simple method for fast and cheap thermal analysis on
  supercooled glass-forming liquids. This ``Thermalization
  Calorimetry'' technique is based on monitoring the temperature and
  its rate of change during heating or cooling of a sample for which
  the thermal power input comes from heat conduction through an
  insulating material, i.e., is proportional to the temperature
  difference between sample and surroundings. The monitored signal
  reflects the sample's specific heat and is sensitive to exo- and
  endothermic processes. The technique is useful for studying
  supercooled liquids and their crystallization, e.g., for locating
  the glass transition and melting point(s), as well as for
  investigating the stability against crystallization and estimating
  the relative change in specific heat between the solid and liquid
  phases at the glass transition.
\end{abstract}

\pacs{}% insert suggested PACS numbers in braces on next line
\keywords{Thermalization, calorimetry, supercooled liquid, glass
  transition, crystallization}

\maketitle %\maketitle must follow title, authors, abstract and \pacs

\section{Introduction}
\label{sec:introduction}
Einstein is often quoted for stating that if you have just one choice
of measurement, go for the specific heat. The point is that this
quantity by integration determines the entropy, the free energy, etc,
i.e., all of the relevant thermodynamics. Indeed, calorimetry is a
standard method throughout the scientific, technical, and industrial
world, which has been commercially available in many different
versions for more than 50 years
\cite{Armstrong1964,Privalov1986,Gmelin1997,Schick2002,Wunderlich2005,READING2006}. Modern
calorimetry is highly sophisticated and measures with unprecedented
accuracy and, even more importantly, on smaller and smaller samples
\cite{Zhuravlev2010,Zhao2015,ThActa}. Thus nanocalorimetry is now
becoming routinely available, which is essential for ultrafast
measurements and expensive samples, e.g., in modern molecular biology
or advanced material science \cite{Mathot2011,ThActa,Zhao2015}.  At
the same time there is, however, a need for ``quick and dirty''
calorimeters in some branches of physics and chemistry in which fast
answers for initial investigations are more important than accuracy.

Our group \cite{GogT} has for many years conducted research into
viscous liquids and the glass transition. ``What is the glass
transition temperature?''\ is the first question one asks when
investigating a new liquid. Here accuracy is not important, but speed,
flexibility, and simplicity are. This paper presents our simple
``Thermalization Calorimetry'' (\textit{TC}) method, which has proven
to be a useful tool for preliminary investigations of glass-forming
liquids.

The TC method is based on recording the temperature as a function of
time for a system that is slowly cooled or heated with a heat
current that depends linearly on the temperature of the sample because
it derives from heat conduction to (or from) the surroundings. The TC technique
is simpler to implement than traditional differential scanning
calorimetry (DSC) because the linearity between heat current and
temperature is obtained by passive insulation of the sample and because no
heating device is required.

In TC changes in specific heat, exothermic, and endothermic processes,
are observed by plotting the rate at which the temperature changes
with time, $\frac{\mathrm{d} T(t)}{\mathrm{d} t}$, versus the
temperature, $T(t)$. Such a plot reveals a wealth of thermodynamic
information about the sample, its glass transition, crystallization,
etc. The TC method is inspired by simple experiments conducted more
than 40 years ago by our colleague J.\ H\o{}jgaard Jensen
\cite{Jensen1972} for demonstrating the glass transition in
propanols. Since then the TC method (nicknamed ``\textit{The Red
  Box}'') has been a workhorse for years in the {\it Glass and Time}
group. TC is our standard tool for initial studies of new
glass-forming liquids by which we locate the glass-transition
temperature and investigate the liquid's stability against
crystallization. Furthermore, TC has turned out to be excellent for
teaching purposes. It has been used for numerous student projects at
all levels from high school to master's project, in which the method's
simplicity and ease of adaptation has allowed for studies of samples
ranging from amorphous drugs to candy and amber.

The paper is structured as follows: Section
\ref{sec:heate-rate-technique} introduces the technique and how to
interpret TC data. Examples of applications to glass transition and
crystallization are given in
Sec.\ \ref{sec:example-application}. Finally, Sec.\ \ref{sec:discussion}
presents a short discussion of pros and cons of the technique and
perspectives on adaptation to further applications.  The appendices
give details on the experimental protocols (Appendix
\ref{sec:exper-prot}), on the technical implementation of the
technique (Appendix \ref{sec:details-electronics}), as well as a
discussion of a simple Tool-type \cite{Tool1946} model of the glass
transition that is useful also for teaching purposes, (Appendix
\ref{sec:simp-model-underst}).

\section{The Thermalization Calorimetry (TC) technique}
\label{sec:heate-rate-technique}

\begin{figure}
  \includegraphics[]{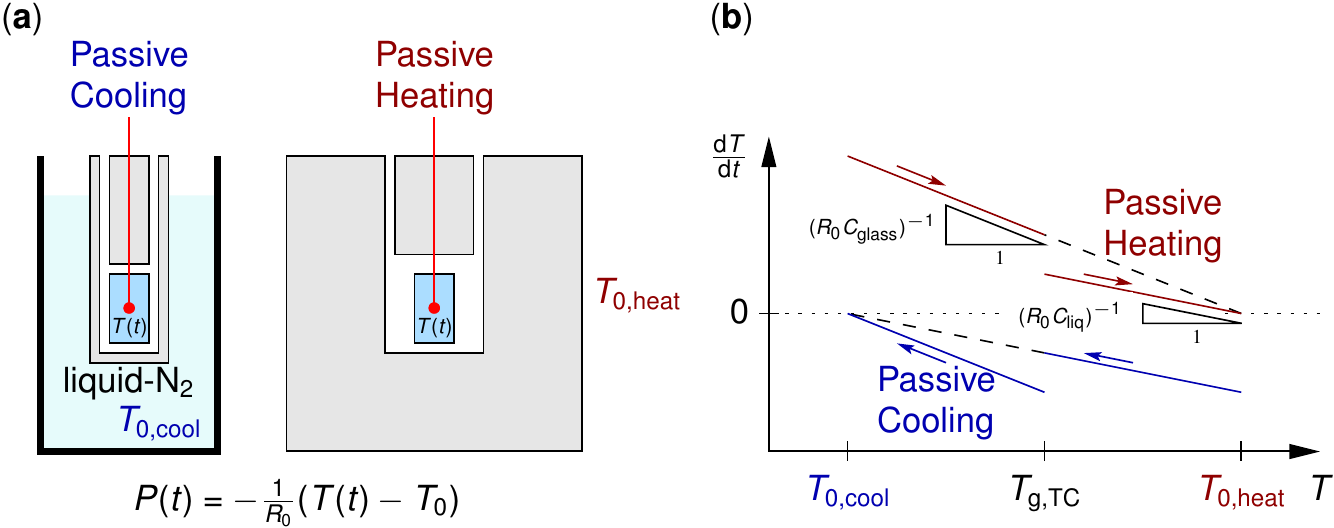}
  \caption{Illustration of the two typical modes of operation of the
    TC technique and the expected \TdTdt{}-traces for a glass forming liquid. \\
    \textbf{a)} The experimental setup. The gray region is insulating
    material, the red line and dot illustrate the thermocouple
    inserted in the sample (marked by blue). The basic assumption is
    that the thermal power into the sample, $P(t)$, is proportional to
    the difference between the temperature of the surroundings, $T_0$,
    and the sample temperature, $T$, with proportional constant
    $1/R_0$ in which $R_0$ is the thermal resistance of the insulating
    material. \textbf{Left} part illustrates slow passive cooling of a
    sample placed in an insulating container inserted directly into
    the cooling agent (which is normally liquid nitrogen). $T_0$ is
    here the temperature of the cooling agent, i.e.,
    \SI{77}{\kelvin}. \textbf{Right} part illustrates a slow passive
    heating of a pre-cooled glassy sample using a block of insulation
    material. $T_0$ is here the (room)
    temperature outside the insulating material.  \\
    \textbf{b)} Expected \TdTdt{}-traces for the two protocols applied
    to a glass-forming liquid with constant specific heat in the
    liquid ($C_\text{liq})$ and the glass $(C_\text{glass})$. The
    curves change slope and show a ``jump'' at the glass transition
    temperature, $T_{\text{g},\text{TC}}$, due to the change in
    specific heat between the equilibrium supercooled liquid and the
    glass.  The dashed lines are guides to the eye, extending to
    $(T_0,0)$ in the both cases. See Sec.\ \ref{sec:interpretation}
    for details on the interpretation. }
  \label{fig:expprotocol}
\end{figure}

Figure \ref{fig:expprotocol} illustrates the two ways we normally use
the TC technique. As mentioned in the Introduction, the TC method is
based on heating or cooling via a passive, insulating material; this
results in a heat current that is proportional to the temperature
difference between sample and surroundings. During this process the
temperature and its time derivative are recorded. In the following we
briefly describe the TC technique and the two protocols normally used for
studying glass-forming molecular liquids. The technique can be
adapted to other conditions and protocols, some of which are briefly 
discussed in Sec.\ \ref{sec:discussion}.

A convenient and commonly used TC protocol for characterizing glass-forming
molecular liquids is a ``fast quench'' to the glassy state followed by
the ``slow heating'' of thermalization. Alternatively, the liquids may
also be studied under ``slow cooling'' and subsequent ``slow
heating'' (examples of all combinations are given in
Sec.\ \ref{sec:glass-transition}). Figure \ref{fig:expprotocol}
illustrates the ``slow heating'' and ``slow cooling''
arrangements. For details on the experimental protocols associated
with these experiments, see Appendix \ref{sec:exper-prot}.

In all experiments the liquid is positioned in a small glass test tube
(\SI{8}{\milli\meter} in diameter and \SI{40}{\milli\meter} in length)
and a thermocouple-based thermometer is submerged in the
sample. Li\-quid nitrogen provides an excellent cooling agent for
studying most molecular glass-forming liquids, and room-temperature is
likewise a convenient target temperature for heating experiments. A
fast quench is performed by dipping the test tube into a jar of
boiling liquid nitrogen. For the slow heating protocol, from liquid
nitrogen temperature to room temperature, the pre-cooled sample is
subsequently quickly transferred to a block of expanded polystyrene
(EPS) --- pre-cooled to establish the temperature gradient --- with
$\approx \SI{5}{\centi\meter}$ wall thickness. This combination of
sample size and insulation block gives typical initial heating rates
of a few tenths of a Kelvin per second ($\approx
10\si{\kelvin/\minute}$). For ``slow cooling'' similar cooling rates
can be obtained by insulating the test tube from the liquid nitrogen
by a few millimeters of insulating material. The temperature range and
cooling/heating rates can easily be modified by using other
cooling/heating agents and insulating arrangements.

Two different implementation of the TC technique exist in our lab: the
original version based on analog electronics for differentiation and
averaging of the data (used for the majority of data presented in this
paper) and a more recent digital version in which the differentiation
is performed after digitization of the $T(t)$ signal. The
implementations of both versions are discussed in Appendix
\ref{sec:details-electronics}.

\subsection*{Interpretation of \TdTdt{}-graphs}
\label{sec:interpretation}
The data analysis of the TC technique is based on plotting
\TdTdt{}-graphs. The basis of the interpretation is that the rate at
which the sample changes its temperature, $\diff{T(t)}{t}$, is
inversely proportional to the specific heat of the sample, $C$, at a
given heat current input, $P(t)=\diff{Q}{t}$. The onset is the
definition of the specific heat:
\begin{align}
  \label{eq:Cdeff}
  \Delta T &= \frac{1}{C} Q
\end{align}
where $\Delta T$ is the temperature increase after introducing some
amount of heat, $Q$. Taking the time derivative leads to the desired relation:
\begin{align}\label{eq:dTdt}
  \diff{T(t)}{t} &= \frac{1}{C}\diff{Q}{t}=\frac{1}{C} P(t).
\end{align}
In the long time limit this holds for the macroscopic sample
investigated here (neglecting heat diffusion in the sample).

A fundamental assumption of the analysis is that the heat current into
the sample is proportional to the difference between the target
temperature, $T_0$, and the sample temperature:
\begin{align}
  \label{eq:power}
  P(t)=-\frac{1}{R_0}(T(t)-T_0)
\end{align}
where $R_0$ is the thermal resistance in the experiment (assumed to be temperature independent).

Combining the above equations (Eq.\ \ref{eq:dTdt} and \ref{eq:power})
leads to the following differential equation:
\begin{align}
  \label{eq:basic_eq}
  \diff{T(t)}{t} &= -\frac{1}{R_0C} (T(t)-T_0).
\end{align}
This is easily solved analytically for constant $R_0C$, 
\cite{footnote1}
but that is \textit{not} the route we take for analyzing data.
Instead, direct information is obtained by plotting $\diff{T}{t}$ as a
function of $T$ (see Fig.\ \ref{fig:expprotocol}). For constant $C$
the graph is a straight line with slope $-\frac{1}{R_0C}$ ending at the point
$(T,\diff{T}{t})=(T_0,0)$.  The instantaneous value of
$-\frac{1}{R_0C}$ at a given \TdTdt{} point can be found as the slope of
the secant from that point to $(T_0,0)$.

The ability to directly monitor changes in specific heat as a function
of temperature is particularly useful for investigating glass-forming
liquids, because a signature of the glass transition is a change in
specific heat, as illustrated on Fig.\ \ref{fig:expprotocol}. This
change of specific heat is the result of the ``freezing in'' of the
structural degrees of freedom when the liquid falls out of equilibrium
to form a glass \cite{Kauzmann1948}. The glass transition has the
appearance of a second-order phase transition in the Ehrenfest sense,
but as is well known the transition is a dynamic phenomenon and the
transition temperature changes (logarithmically) with cooling rate
\cite{Dyre2006}.

Exothermic and endothermic processes likewise give a very clear TC
signal because they dramatically change the overall \TdTdt{}-trace by,
respectively, a large increase in $\diff{T}{t}$ and 
$\diff{T}{t}$ approaching 0.

\section{Applications}
\label{sec:example-application}

\begin{figure}  
  \hbox{\hspace{-0.7cm}\includegraphics{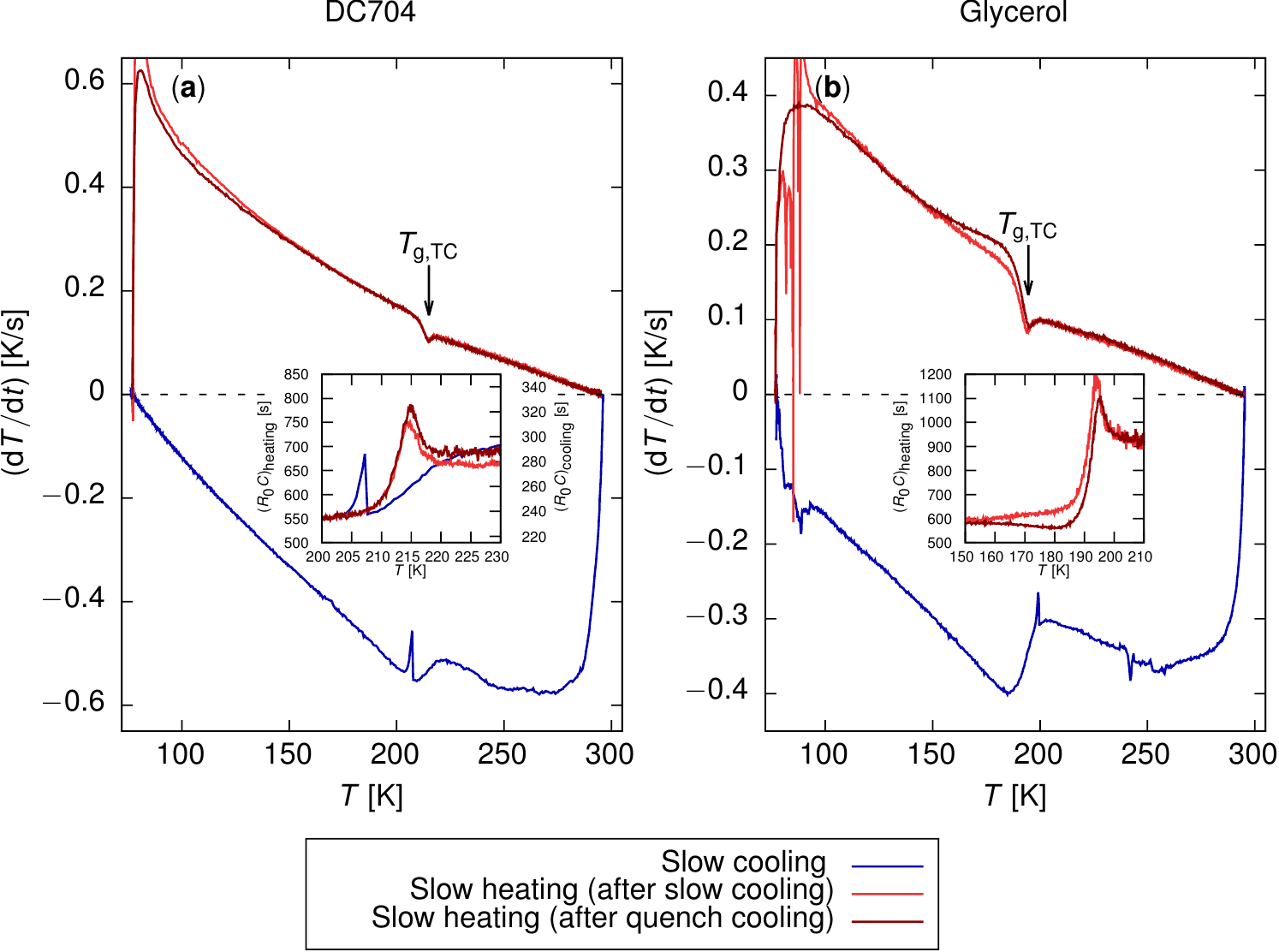}}
  \caption{Glass transition studied by TC. Measurements on two typical
    glass forming liquid, (a) DC704 and (b) glycerol (see Ref.\
    \onlinecite{Sampel} for details on the samples).
    The dark red curves show heating after a fast quench to liquid
    nitrogen temperature. The light red and blue curves show a cycle
    of slow cooling to liquid nitrogen temperature (blue curves)
    followed by slow heating to room temperature (light red
    curves). The glass transition is clearly seen in both cooling and
    heating. The inserts show the apparent specific heat represented
    as the $R_0C$ value from Eq.\ (\ref{eq:basic_eq}) (notice the
    different y-axis for the heating and cooling $R_0C$ curves, due to
    the difference in $R_0$ between the two protocols).}
  \label{fig:measurement}
\end{figure}

The TC technique is useful for studying different aspects of
supercooled liquids. In the following we present examples of how the
technique can be used for locating the glass-transition temperature
and for studying crystallization.

\subsection{Data from two glass-forming liquids}
\label{sec:glass-transition}

Two typical molecular glass-forming liquids,
tetramethyl-tetraphenyl-trisiloxane (commercial name: DC704) and
glycerol, have been investigated \cite{Sampel}. The former has been
studied in great detail by the {\it Glass and Time } group by several
techniques
\cite{Christensen1994,Jakobsen2005,Gundermann2011,Jakobsen2012,Hecksher2013};
the latter is a prototypical glass-forming liquid.

\subsubsection{Measurements}
\label{sec:measurements}
Figure \ref{fig:measurement} shows \TdTdt{}-traces from measurements
on the two liquids. The figure includes data for ``slow heating
after quench cooling'' and temperature cycles consisting of a ``slow
cooling'' followed by a ``slow heating''.  The heating curves start
in thermal equilibrium at liquid nitrogen temperature
($\SI{77.3}{\kelvin}$). Initially they show a sharp increase in
$\frac{\mathrm{d} T}{\mathrm{d} t}$ deriving from the transfer of the
sample to the insulating block. A small $\frac{\mathrm{d}
  T}{\mathrm{d} t}$ overshoot is often observed, depending on how fast
this transfer is performed and how well the insulating block has been
pre-equilibrated (see Appendix \ref{sec:exper-prot}). The trace
quickly becomes approximately linear in time. The cooling curves
likewise start in equilibrium at room temperature with a fast decrease
in $\frac{\mathrm{d} T}{\mathrm{d} t}$ until equilibrium cooling
conditions have been established.

A clear signature of the glass transition (at $\SI{215}{\kelvin}$ for
DC704 and \SIrange{194}{195}{\kelvin} for glycerol) is seen during
both heating and cooling as changes in the slope of the \TdTdt{}-trace
(compare Fig.\ \ref{fig:expprotocol}). Note that the signature
of the glass transition is very robust, also when changing cooling
protocol, and that the cooling and heating curves are consistent. 

Small sharp peaks close to the glass transition often appear on the
cooling curves, as seen at $\approx \SI{205}{\kelvin}$ for DC704. We
believe that these stem from crack formation in the sample as it
solidifies. The cooling curve for glycerol display a problem sometimes
encountered in cooling. The \TdTdt{}-trace on the low
temperature side of the glass-transition does not ``point to'' the
$(T_0,0)$ point. We attribute this to the rather primitive insulating
arrangement used; if cooling curves are needed for accurately
evaluating the relative change in specific heat (discussed below),
a better arrangement can be constructed; however, the identified
glass-transition temperature is not affected by this.

\subsubsection{Analysis}
\label{sec:analysis}
Even though the glass transition is well resolved in the
\TdTdt{}-traces, it covers a non-negligible temperature range. In order
to derive an experimentally well-defined number we define the
experimental TC glass-transition temperature,
$T_{\mathrm{g},\text{TC}}$, to be the temperature of the local minimum
on the heating curve.

The $T_{\mathrm{g},\text{TC}}$ values are given in Table \ref{tab:res}
together with \Tg{} and characteristic times at
$T=T_{\mathrm{g},\text{TC}}$ from
studies\cite{Jakobsen2012,SanzGlycerol} of the equilibrium shear
modulus ($\tau_G$), dielectric ($\tau_\epsilon$), and specific heat
($\tau_{C_l}$). It is seen that the $T_{\mathrm{g},\text{TC}}$ is at
slightly higher temperature than the traditionally defined \Tg{} of
$\tau=\SI{100}{\second}$. In fact, the glass-transition temperature is
not uniquely defined; different experimental methods has different
characteristic time scales at the same temperature --- therefore
a slightly different \Tg{}.

The apparent specific heat, represented as $R_0C$, can be found from
the data using Eq.\ (\ref{eq:basic_eq}), and is shown in the insets of
Fig.\ \ref{fig:measurement}. The change in specific heat between the
liquid and glass is clear in this representation. The above discussed
experimentally determined $T_{\mathrm{g},\text{TC}}$ corresponds to
the maximum of the observed overshoot in the specific heat. From the
$R_0C$ data an estimate of the relative change in specific heat
between glass and liquid can be found; these data are given in
Table \ref{tab:res} together with literature data. It is seen that the
TC technique provides a good first estimate of this quantity. For
DC704 the cooling and heating curves give consistent results. This
could not be checked for glycerol due to the problems with the
\TdTdt{}-trace endpoint for the cooling measurement discussed in
Sec.\ \ref{sec:measurements}. 

A significant difference between the heating curves of quenched and
slowly cooled glycerol is observed. This difference can be
rationalized by the difference in thermal history and hence the
glasses formed. However, depending on the properties of the liquid
studied the signature of thermal history will be more or less visible,
as is seen when comparing the DC704 and glycerol data. A simple model
that captures the essence of glass formation, including the
differences between fast and slow cooling, is presented in  Appendix
\ref{sec:simp-model-underst}.
 
\setlength{\extrarowheight}{5pt}%
\begin{table*}
  \centering
  \begin{tabular}{@{\extracolsep{7pt}}lrrrrrrrrr}
     & \multicolumn{4}{c}{$T_\text{g}$}   
     &\multicolumn{3}{c}{$\tau$ at $T_\text{g,TC}$}
     &\multicolumn{2}{c}{\raisebox{0.3 em}{$\frac{\Delta C}{C_\text{liq}}$}}
\\
    \cline{2-5} \cline{6-8} \cline{9-10}
    & $T_{\text{g},\text{TC}}$ & $T_{\text{g},{C_l}}$ & $T_{\text{g},\epsilon}$ & $T_{\text{g},G}$  
   & $\tau_{C_l}$ & $\tau_{\epsilon}$ & $\tau_G$ 
   & $\text{TC}$ & Lit.\    
\\ \hline
\multicolumn{1}{l|}{DC704}
& \SI{215}{\kelvin} & \SI{211.5}{\kelvin}&\SI{210}{\kelvin}&\SI{208}{\kelvin} 
& \SI{4.8}{\second}&  \SI{1.4}{\second}&\SI{0.3}{\second}
& \num{0.2} &  0.2 [\onlinecite{Gundermann2011}] 
\\
\multicolumn{1}{l|}{Glycerol}
& \SIrange{194}{195}{\kelvin} &&  \SI{186}{\kelvin} & \SI{180}{\kelvin}
& & \SI{1.9}{\second} &\SI{0.03}{\second} 
& \num{0.3} & \numrange{0.4}{0.5} [\onlinecite{Christensen1984,*Birge1985,*Yamamuro1998,*Donth2001,*Andersson2016}]
  \end{tabular}
  \caption[]{Glass transition temperature ($T_\mathrm{g}$) from TC and
    three dynamic response functions (shear modulus, 
    ${G}(\omega)$, dielectric constant, ${\epsilon}(\omega)$, and
    specific heat, ${C}_{l}(\omega)$),  
    characteristic timescale ($\tau$) of the same three response functions at $T_\text{g,TC}$,
    and  relative change in specific heat between glass and liquid ($\Delta
    C/C_\text{liq}=({C_\text{liq}-C_\text{glass}})/{C_\text{liq}}$).    
    The values are based on the TC data shown in Fig.\
    \ref{fig:measurement} and from
    measurements\cite{SanzGlycerol,Jakobsen2012} of the three 
    dynamic quantities on the thermo-viscoelastic 
    (metastable) equilibrium viscous liquid. 
    $T_\text{g}$ from TC is found as described in the main text, from
    the local minimum on the heating curve, whereas for the
    three response functions $T_\mathrm{g}$ is defined as the
    temperature at which the characteristic time is
    \SI{100}{\second}. \cite{footnote5}
    $\tau$ at $T_\text{g,TC}$ is found from the same dynamical data
    for the three response functions. 
    $\Delta C/C_\text{liq}$ from TC is found from the data shown on
    the insert of Fig.\ \ref{fig:measurement} and is compared to
    values from literature.
  }
  \label{tab:res}
\end{table*} 

\subsection{Glass-transition temperature of glycerol-water mixtures}
\label{sec:GlycerolWaterMix}
Liquid-liquid mixtures in general and alcohol-water solutions in
particular have been widely explored, not only in fundamental studies,
but also for their potential applicability \cite{sato}. Here we take
as a case study a family of glycerol-water mixtures. Glycerol-water is
a well-known miscible system in which the crystallization ability of
water is highly suppressed. For this reason, many authors have used
this mixture to address fundamental questions of the physics of pure
liquid water\cite{Murata2012,Amann2016}. Glycerol-water mixtures have also
received much interest due to their utilization as bioprotective
agents of cells and proteins \cite{Salt1961,davis}.

\begin{figure}
  \centering
    \hbox{\hspace{-0.7cm}\includegraphics{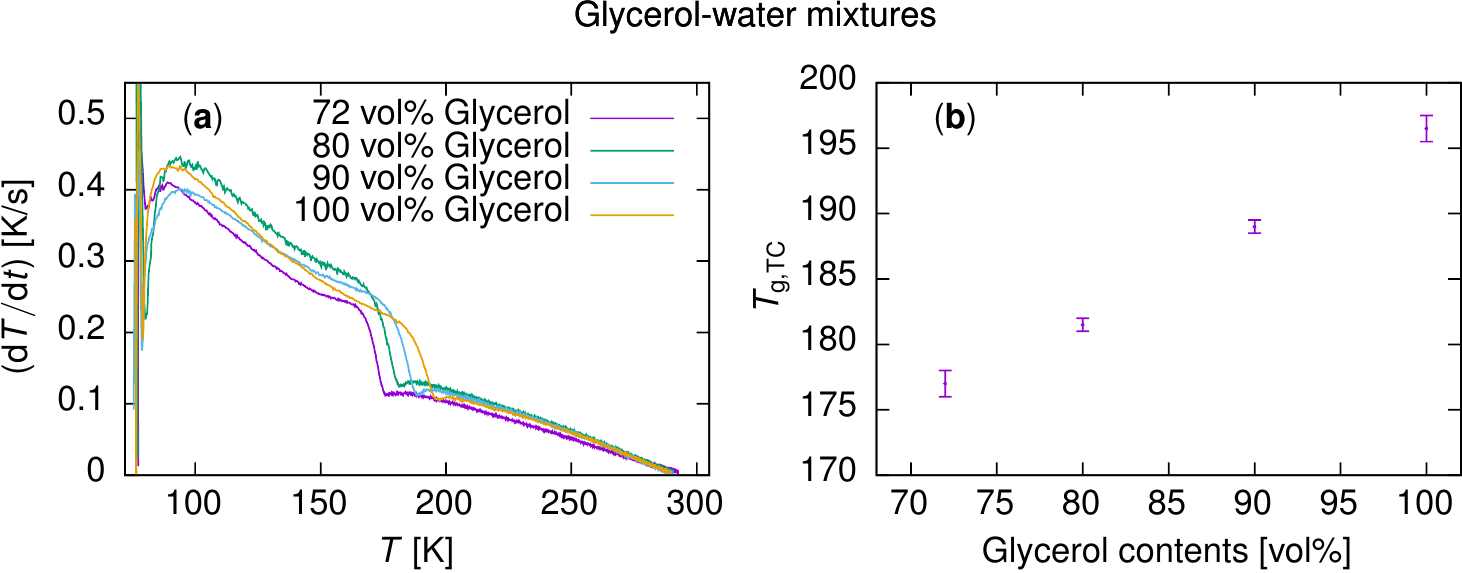}}
  \caption{Glass-transition temperature of glycerol-water mixtures
    studied by the TC technique. \textbf{(a)} \TdTdt-traces for
    glycerol-water mixtures with varying glycerol contents.  \textbf{(b)}
    Glass-transition temperature, $T_{\text{g},\text{TC}}$, as a
    function of glycerol content given as the volume fraction of
    glycerol. Error bars are estimated from the shape of the local
    minimum in the \TdTdt{}-trace defining \TgTC{}.  }
  \label{fig:G_W_mix}
\end{figure}

Four glycerol-water mixtures (72, 80, 90, \SI{100}{\percent} volume
fraction of glycerol) were prepared by mixing anhydrous glycerol
(purity $\geq$ \SI{99.5}{\percent}) with distilled and deionized pure
water (Arium 611$^{\circledR}$). Estimates of the glass-transition
temperature of pure water give values lower than those corresponding
to pure glycerol (the exact glass-transition temperature of water is
not known as water crystallizes by spontaneous homogeneous
crystallization in the supercooled regime before reaching the glass
transition, see, e.g., Ref.\ \onlinecite{Debenedetti2003}). One
expects that miscible mixtures of glycerol-water are characterized by
a monotonic decrease of \Tg{} when the water content increases. Figure
\ref{fig:G_W_mix}(a) shows the \TdTdt{}-traces from slow heating of
initially quenched glycerol-water mixtures.  The signature of the
glass transition shifts towards lower temperatures with the water
content, as expected. Figure \ref{fig:G_W_mix}(b) shows how the values
of \TgTC{} become lower when the volume fraction of glycerol decreases.

\subsection{Crystallization and melting}
\label{sec:Crystallization}

\begin{figure}
    \hbox{\hspace{-0.7cm}\includegraphics{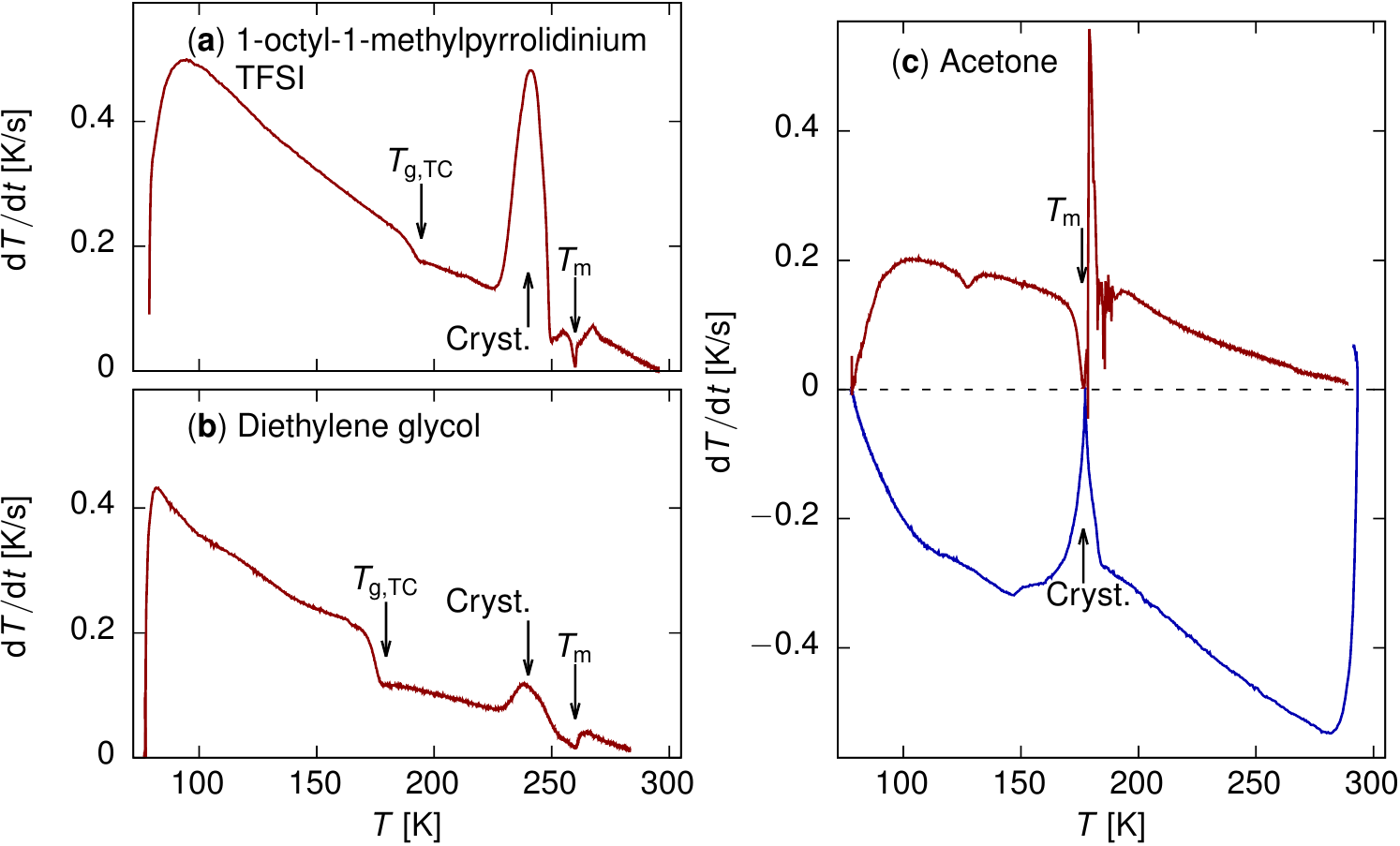}}
  \caption{Crystallization, melting, and glass transition studied on
    three samples \cite{Sampel} by the TC technique (melting and
    glass-transition temperatures are given in Table \ref{tab:cryst}).
    \textbf{(a)} and \textbf{(b)} \TdTdt-trace for slow heating after
    fast quench showing glass transition, crystallization from the
    supercooled liquid, and melting.  \textbf{(c)} \TdTdt-trace for
    slow cooling and subsequent slow heating, showing crystallization
    in cooling and melting in heating. See Sec.\
    \ref{sec:Crystallization} for a discussion of further features.}
  \label{fig:crystalization}
\end{figure}

\begin{table*}
  \centering
  \begin{tabular}{lrlr}& $T_{\text{m}}$ && $T_{\text{g},\text{TC}}$  \\ \hline
    \multicolumn{1}{l|}{1-Octyl-1-methylpyrrolidinium   TFSI}  &\SI{260}{\kelvin} &&
    \SIrange{194}{195}{\kelvin} \\
    \multicolumn{1}{l|}{Diethylene glycol} & \SI{260}{\kelvin} & (\SI{262.7}{\kelvin} [\onlinecite{CRC}]) &
    \SIrange{178}{181}{\kelvin}\\
    \multicolumn{1}{l|}{Acetone} &  \SIrange{176.5}{177.5}{\kelvin} & (\SI{178.8}{\kelvin} [\onlinecite{Ibberson1995}])  \\
  \end{tabular}
  \caption{Melting, $T_{\text{m}}$, and glass-transition temperature,
    $T_{\text{g},\text{TC}}$, determined from TC on three substances
    \cite{Sampel} (raw TC data shown on Fig.\ \ref{fig:crystalization}). Numbers in parenthesis are from the literature. }
  \label{tab:cryst}
\end{table*}

The TC technique is well suited
for studying the stability of a supercooled liquid against
crystallization and locating the melting point of the crystal.

In pure systems, at sufficiently low temperatures the
thermodynamically stable form is crystalline. When liquids are
supercooled and situated between the melting temperature (\Tm) and the
glass-transition temperature (\Tg), these are susceptible to
crystallization. This transition of a metastable liquid into a crystal
is controlled by several factors like the rate of nucleation and
growth of the crystallites, viscosity, and the heating or cooling rate
to which the system is subjected.  Most liquids are easily supercooled
to form a glass when cooled fast enough. Some systems, however,
recrystallize during heating at similar rates from the glassy
state. The transformation of supercooled liquids into crystals (known
as ``cold crystallization'') is of great interest in fundamental
science as well as in industry \cite{karolina}.

Data on the following three liquids with different crystallization
behavior is presented in the following \cite{Sampel}:
1-octyl-1-methylpyrrolidinium TFSI (OctPyr-TFSI) (a room-temperature
ionic liquid), diethylene glycol (a typical molecular glass-forming
liquid), and acetone (a small-molecule solvent). Figure
\ref{fig:crystalization} shows TC data for the crystallization of
these liquids and Table \ref{tab:cryst} summarizes the melting and
glass transition temperatures derived from TC.

Figure \ref{fig:crystalization}(a) shows a heating curve for
OctPyr-TFSI initially quenched in liquid nitrogen. Going
from low to high temperature the glass transition is seen at
\SIrange{194}{195}{\kelvin}, followed by crystallization of the
supercooled liquid around \SI{230}{\kelvin} (which is, of course, an
exothermic process as seen in the large increase in
$\diff{T}{t}$). Finally, the melting of the crystal is observed at
 as an endothermic process ($T_\text{m}=260$ K), seen as the sharp dip in
$\diff{T}{t}$ down to 0. 

In heating, an exothermic process leads to a rather wide peak
  in the \TdTdt{}-trace, as the process increases the heating rate
  ($\diff{T}{t}$) over the baseline behavior until all excess heat is
  released, and simultaneously increases $T$. Conversely, endothermic
  process leads to a sharp peak, as $T$ is constant and
  $\diff{T}{t}=0$ until enough heat has been provided to the sample. 

Figure \ref{fig:crystalization}(b) shows a similar curve for
diethylene glycol, which has a more pronounced glass transition, but a
much smaller exothermic signal form the crystallization and a not so
well-defined melting. Although this is the typical behavior for molecular
liquids, the TC technique gives a good estimate of the
melting point and a simple test of the stability of the supercooled
liquid against crystallization.

As a final example of a crystallizing liquid data on acetone are
presented in Fig.\ \ref{fig:crystalization}(c), showing \TdTdt-traces
for a slow cooling and subsequent slow heating. For most molecular
liquids, crystallization is suppressed during cooling, even at
moderately slow cooling rates (the rate used here is $\approx
\SI{0.5}{\kelvin/\second}$), but for acetone this is not the case,
since acetone unavoidably crystallizes around its freezing point at
the cooling rates TC realizes.  On cooling we thus observe an
exothermic transition around \SI{177}{\kelvin} (blue curve) that
corresponds to the crystallization point of pure acetone.  On slow
heating of the crystal (red curve) we observe an endothermic peak
around \SI{177}{\kelvin} that confirms the melting of crystalline
phase of acetone. At slightly higher temperatures a narrow exothermic
peak emerges followed by a region of small peaks. The origin of this
is unclear, but the observation resembles additional crystallization
and melting events.  A weak signal is observed on heating (even more
weak on cooling) at \SI{130}{\kelvin}, which appears like a glass
transition, however, in the crystalline phase. Acetone is known to
have low-temperature thermal transitions \cite{Kelley1929}, the origin
of which has been debated during the last decades (see, e.g.,\ Ref.\
\onlinecite{Ibberson1995,Shin2014}). It has been proposed that an
order-disorder transition or more complex phase transitions exists
\cite{Kelley1929,Ibberson1995}, but it has also been argued that
variation in the strength of the electrostatic interactions between
polar groups along the crystalline lattice cause these transitions
\cite{Allan1999}. Dielectric spectroscopy on the crystal shows that a
dielectric process exists in the crystalline phase \cite{Sanz2005},
showing that some dynamical processes are still active in the crystal,
which indicates a plastic crystalline behavior. Altogether, this
demonstrates that TC also can be used for ``quick and dirty''
investigations of complex phenomena.

Stability against crystallization is a complicated problem, and the
above sketched experiments are by no means meant to represent all
different ways crystallization can occur. Due to the
simplicity of the TC setup, different types of crystallization
experiments can be performed \textit{ad hoc}, e.g., by keeping the
liquid at isothermal conditions above \Tg{} and studying whether
excess heat is generated by crystallization in a subsequent heating.

\section{Summary}
\label{sec:discussion}

We have presented the general principle of thermalization calorimetry
(TC), a simple technique for investigating supercooling, glass
transition, and crystallization of liquids. The key idea is to monitor
\TdTdt{}-curves for a system in which the thermal input power is
proportional to the temperature difference to the sample
surroundings. Changes in specific heat, glass transition, exothermic,
and endothermic processes all have a clear signal in the
\TdTdt{}-curves.

Uses of the TC technique for locating the glass-transition
temperature, studying stability against crystallization, determining
the melting point, and studying complex crystallization behavior have
been illustrated. The TC technique should not be seen as a way to
determine \Tg{} and \Tm{} precisely, which in any case is not possible
for \Tg{} (see table \ref{tab:res} and Ref.\
\onlinecite{Jakobsen2012}), but rather as a quick way of getting a
good estimate of such quantities as well as, e.g., ratios of specific
heat between solid and liquid.

An advantage of the TC technique is that it can easily be adopted to
different experimental situations. By changing the temperature of the
environment and the insulation used, different temperature regimes and
different cooling-rate regimes can be explored. A straightforward modification of the TC technique is
to use an oven as outer temperature, in this way allowing for studying
substances with high \Tg{} and \Tm{} applicable, e.g., for many substances used in
food science (as sugar and cocoa butter). By changing the thermometer
type the TC technique can also be adopted to lower temperatures where
thermocouples are less efficient.

TC can be realized in a variety of ways; all that is needed is a
recording of $T$ and $\diff{T}{t}$. Initially, we used analog
electronics for the computations, with the later addition of a
build-in ADC (Analog to Digital Converter) for recording the
signal. With the development of good and cheap high resolution ADCs
and small microcontrollers it is now possible to make a fully
digital version with a high degree of flexibility, which can easily be
built, e.g., for educational purposes.
In summary, the TC technique is useful for initial studies of
glass-forming liquids, as a test of new liquid, as well as for teaching
purposes for instance in student projects.

\begin{acknowledgments}
  All students and staff at IMFUFA who have worked with and improved
  the TC-technique over the last 40 years, are gratefully acknowledged
  for helping in making the TC-technique a standard (workhorse)
  technique in our lab.  
  This work was supported by the Danish National Research Foundation via grant
  DNRF61.
\end{acknowledgments}

\appendix

\section{Experimental details}
\label{sec:exper-prot}
This Appendix provides details on the experimental protocols used for
TC measurements; sources of uncertainty are also discussed.

\begin{figure}
  \centering
  \includegraphics[width=0.7\columnwidth]{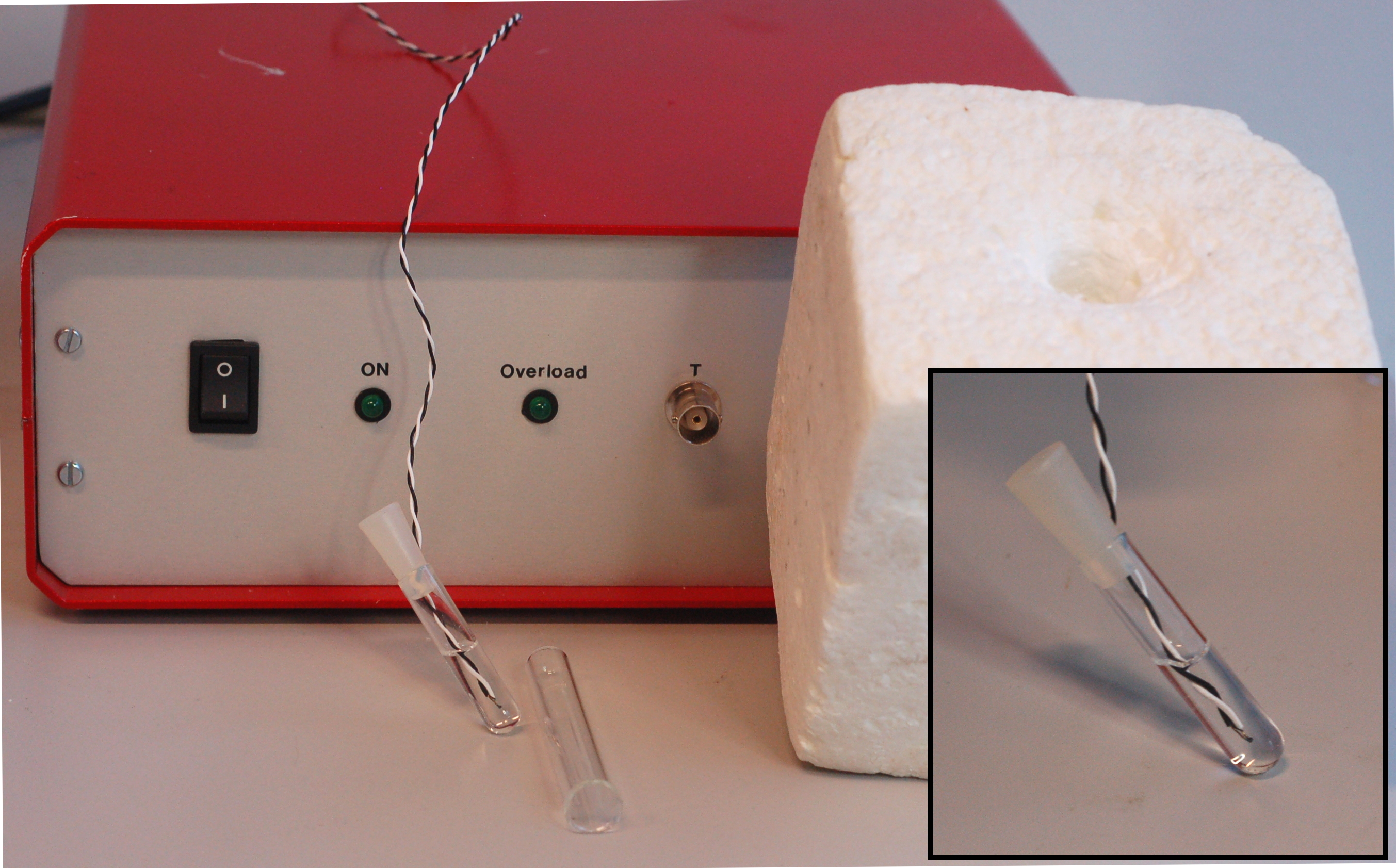}
  \caption{Photo of the TC setup. Shown are the sample with thermometer
    (the thermocouple) (the inset), the
    insulation block of expanded polystyrene used for slow heating,
    the test tube, which filled with liquid nitrogen is inserted into the
    insulation block before the heating experiment for precooling the
    insulation block to establish the temperature gradient. Also,
    the analog TC box (known as the ``Red box'') is seen. The analog
    TC box (Appendix \ref{sec:details-electronics}) normally
    communicates the digitized \TdTdt{} measurements to an attached PC
    over USB. However, an analog representation of the \TdTdt{} signal is
    also available.}
  \label{fig:photo}
\end{figure}

In all experiments the liquid is positioned in a small glass tube
(\SI{8}{\milli\meter} in diameter and \SI{40}{\milli\meter} in
length). The thermocouple is fed through a hole in a soft silicon
plug, which allows for a tight sealing when placed in the beaker (Fig.\ \ref{fig:photo}). 
The thermocouple junction should not touch the beaker and be positioned as
close as possible to the middle of the beaker. Liquid nitrogen is used
as cooling agent. The experiment normally takes place at room
temperature.
For slow heating a block of expanded polystyrene with approximately
$\SI{5}{\centi\meter}$ wall thickness is used as insulating material
(Fig.\ \ref{fig:photo}). With these dimensions initial heating rates of a
few tenths of Kelvin per second ($\approx 10\si{\kelvin/\minute}$) are
typically obtained.

In the slow-cooling protocol the insulation consists of a slightly
larger test tube simply lined with paper, by which cooling rates of
order $\SI{0.5}{\kelvin/\second}$ are obtained. As in the case of
heating, the cooling rate may be varied by changing the insulation
material.

The experimental protocol for slow heating after an initial quench is as
follows:
\begin{enumerate}
\item The sample beaker with liquid is quench cooled directly in
  liquid nitrogen until it has been thermalized.
\item The block of expanded polystyrene is precooled by placing a beaker filled
  with liquid nitrogen in the block for time long enough to establish a
  stationary temperature gradient throughout the insulation block (with
  our dimensions, approximately 5 minutes).
\item The precooling beaker is removed, the sample beaker
  transferred from the liquid nitrogen Dewar to the block of expanded
  polystyrene, and a plug is inserted. This should be done as quickly as
  possible in order to minimize unintended heating of the sample and
  polystyrene block.
\item \TdTdt{} is recorded during the thermalization of the sample.
\end{enumerate}

A number of sources of uncertainties can be identified:
\begin{itemize}
\item The specific heat is in general temperature dependent,
    limiting the temperature ranges over which a straight
    \TdTdt{}-trace can be expected.
\item For slow heating experiments, a too short precooling time or too
  slow transfer of the sample to the block of expanded polystyrene
  lead to distortion of the first part of the heating curve, as Eq.\
  (\ref{eq:power}) will not hold initially.
\item The sample might be rather inhomogeneous after quench
  cooling (it normally fractures), leading to
  inhomogeneous conversion between the phases.
\item Due to the size of the sample significant temperature gradients
  exist.
\item For slow cooling it is often observed that the \TdTdt{}-trace
  does not point to $(T_0,0)$ in the low temperatures regime (as in
  Fig.\ \ref{fig:measurement}(b)). This is attributed to inhomogeneous
  insulation of the sample.
\item The heat capacity of the sample beaker will influence the
  measured specific heat. 
\end{itemize}

\section{Details of our implementation of the TC technique}
\label{sec:details-electronics}
\begin{figure}
     \hbox{\hspace{-0.7cm}\includegraphics[width=1.1\textwidth]{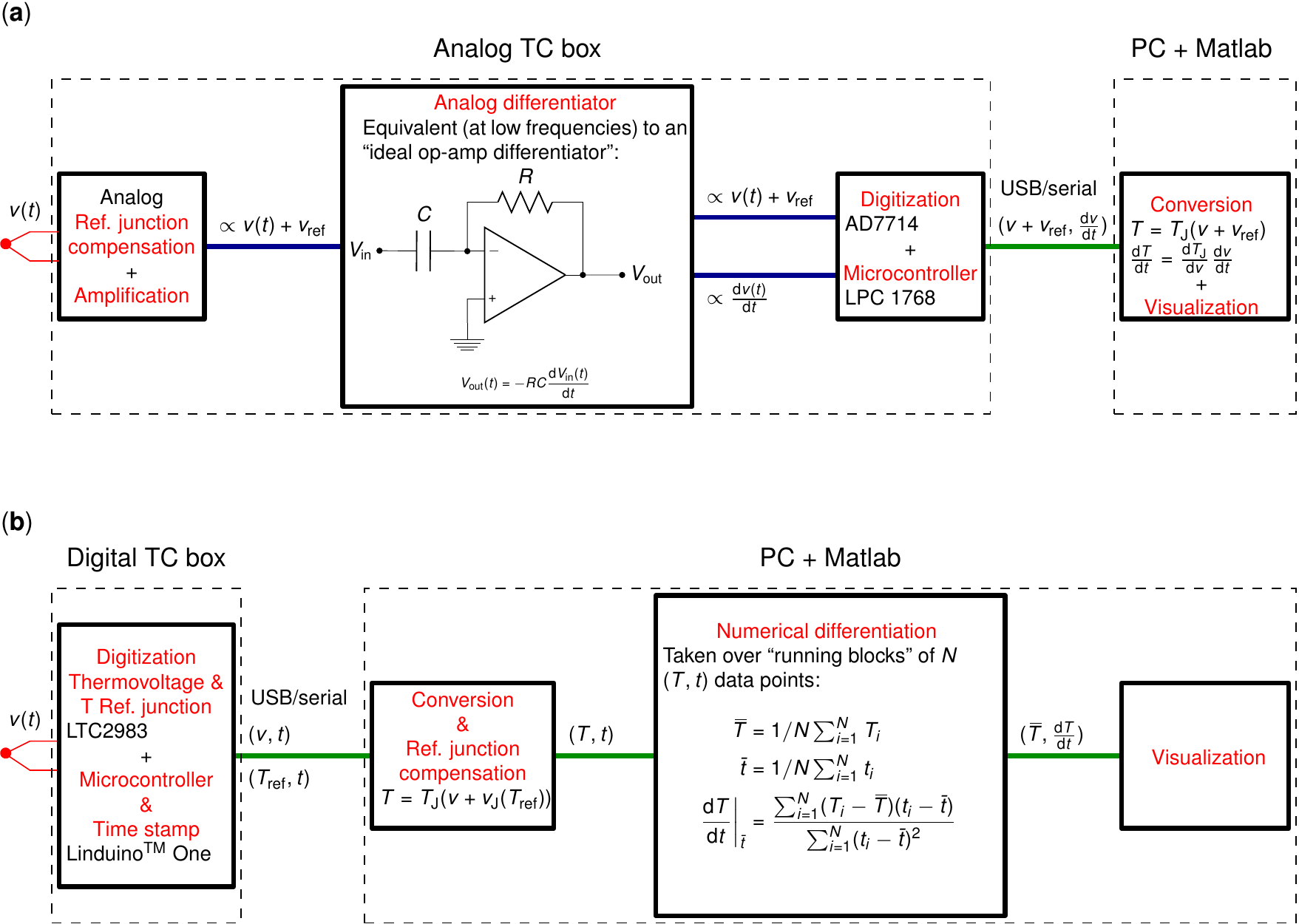}}
   \caption{Schematic representation of two version of the
     TC-technique. $v$ is the thermovoltage,
     $v_\text{ref}$ the thermovoltage associated with the reference
     junction temperature relative to $\SI{0}{\celsius}$,
     $T_\text{J}(v)$ the conversion function\cite{TypeJ} between
     (reference point compensated) thermovoltage and temperature (and
     $v_\text{J}(T)$ the inverse function). Analog quantities are
     designated as time dependent, e.g.,\ $v(t)$, digitized
     quantities as pairs, e.g.,\ $(v,t)$. \textbf{(a)} The original
     analog-based TC-technique implementation consisting of an
     ``Analog TC-box'' (Fig.\ \ref{fig:photo}) which performs
     analog signal processing and differentation of the
     thermovoltage. The analog signals are digitized by a high-quality
     ADC and the results communicated to a PC by a
     microcontroller, where it converted into temperature for
     visualization and analysis. \textbf{(b)} Digital TC-technique
     implementation consisting of a ``Digital TC-box'' (see Appendix \ref{sec:details-electronics}
     for details), which digitizes the thermovoltage and reference
     point temperature and communicates the measurements together with
     time stamps via a microcontroller to the PC. Averaging,
     differentiation, and visualization is done on the
     PC.}
   \label{fig:diagram}
 \end{figure}

The TC technique can be implemented in different ways since the idea
is simply to monitor $T(t)$ with some type of thermometer and record
\TdTdt{}. Two versions of the technique have been developed over the
years at Roskilde University: An analog based and a digital based
version. In the following the two methods are described
briefly\cite{footnote2} and compared (see Fig.\ \ref{fig:diagram} for
a schematic representation of the two implementations).

Common to both version is that a ``J'' type thermocouple is used as
thermometer. It has a reasonable sensitivity in the temperature range
from room temperature ($\SI{50}{\micro\volt/\kelvin}$) to liquid
nitrogen temperature ($\SI{20}{\micro\volt/\kelvin}$). A thermocouple
has the advantage that the heat capacity of the thermometer itself
is very small and that it is mechanically robust and cheap. Both
implementations use a microcontroller for delivering the results to a
PC in which the data are analyzed and visualized using the
Matlab software package.

The analog implementation developed at Roskilde University
approximately 40 years ago is based on analog electronics (Fig.\
\ref{fig:diagram}(a)). It utilizes high-quality operation amplifiers
(op-amp) for analog amplification, differentiation, and low-pass
filtering of the thermocouple signal, and it has analog reference-point
temperature compensation. Originally the analog output signal
proportional to \vdvdt{} (where $v$ is the thermovoltage) was
visualized by means of an ``XY-plotter''. Currently, the analog
signal is digitized using a high resolution analog-to-digital
converter (ADC) connected to a microcontroller that communicates the
result to a PC. In the control and visualization software
the thermovoltage ($v(t)$) and its time derivative 
($\diff{v}{t}$) are converted to actual temperature ($T$) and
temperature rate ($\diff{T}{t}$). The data presented in this paper are
obtained using this version of the TC technique.

The analog version has good resolution and low noise. However, it
requires the ability to build custom-made electronic circuits,
which might be out of range for some applications, e.g.\ for use in
high-school physics education. The digital version takes advantage of
the fact that inexpensive, high-quality microcontrollers and ADC's have
become available in recent years.

Our digital TC-technique implementation uses the LTC2983
``Multi-Sensor High Accuracy Digital Temperature Measurement System''
chip (Linear Technology). The LTC2983 has the advantage that
thermocouples (and other commonly used types of thermometers) can be
directly connected to the chip; the signal is measured by a 
high-quality 24-bit ADC. Furthermore, the LTC2983 has multiple input
channels, allowing for easy digital ``reference junction compensation'' by
having a second thermometer measuring the temperature of the ``reference
junction''.

The LTC2983 is available on a demo board for which ``daughter boards''
for different thermometers (including thermocouples) are available,
which easily connect to a microcontroller\cite{footnote3}.  These
three boards (LCT2983 demo board, thermocouple daughter board, and
microcontroller) provides a full TC system.

The LTC2983 is configured to give the raw digitized thermovoltage
without performing any automated reference junction compensation (this
introduces additional noise in the temperature signal) or using the
internal table for converting thermovoltages to temperature (this was
found not to work well at low temperatures). Including time for
communicating results to the PC, the measuring frequency is $\approx$
\SI{5.8}{\hertz}. The temperature of the ``reference junction'' is
measured every $\approx$ \SI{7}{\second}; for the presented data the
``reference junction'' temperature was to a good approximation
temperature independent. Besides controlling the LTC2983 and
communicating with the PC the microcontroller also time stamps all
measurements.

The \TdTdt{} data are found by calculating mean and linear-regression
slope over ``running blocks'' of $N$ $(t,T(t))$ data points. To obtain
data of comparable quality to those of the analog version of
TC-technique, the averaging and slope calculations is done over
blocks of $N=60$ points corresponding to averaging over $\approx$
\SI{10}{\second}. With this block size the standard derivation of the
slope becomes of order \SIrange[range-units = brackets,scientific-notation = fixed,fixed-exponent = -3]{1E-3}{2E-3}{\kelvin/\second}. Larger $N$
smears our the \TdTdt{}-trace, smaller $N$ introduces more noise.

\begin{figure}
  \centering  
  \includegraphics{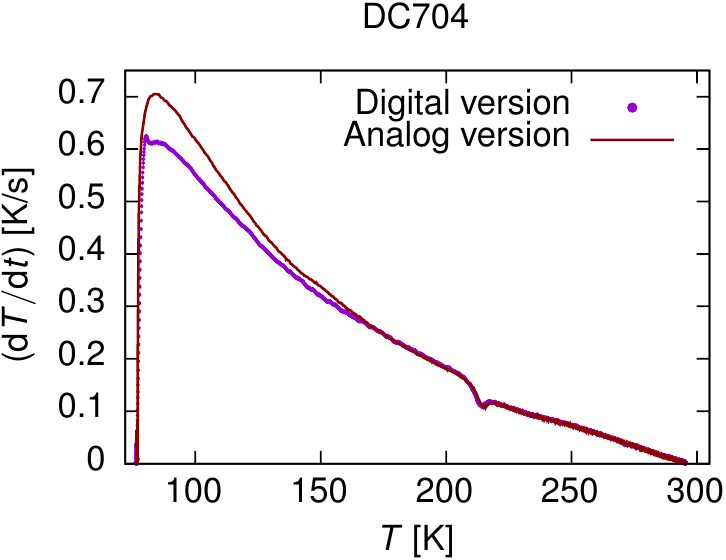}
  \caption{Comparison of the analog and digital versions of the TC
    technique applied to a slow heating of an initially quenched
    sample of the glass-forming liquid DC704. }
  \label{fig:DigitalVsAnalog}
\end{figure}

Figure \ref{fig:DigitalVsAnalog} shows a comparison of the data
obtained by the analog and digital versions of the TC technique,
demonstrating consistent results.

If a shorter averaging time is desired, e.g., for monitoring relatively
fast processes, a simple amplifier can easily be added to the the
digital TC implementation. Using a high-quality op-amp in a
standard inverting amplifier configuration\cite{footnote4}, the
necessary averaging is reduced to $N=20$ points (corresponding to
$\approx \SI{3.3}{\second}$) for noise levels in the slope of order
\SIrange[range-units = brackets,scientific-notation =
fixed,fixed-exponent = -3]{1e-3}{2e-3}{\kelvin/\second}.

\section{A simple model for understanding the TC signal of a
  glass-forming liquid}
\label{sec:simp-model-underst}

\begin{figure}
    \hbox{\hspace{-0.7cm}\includegraphics{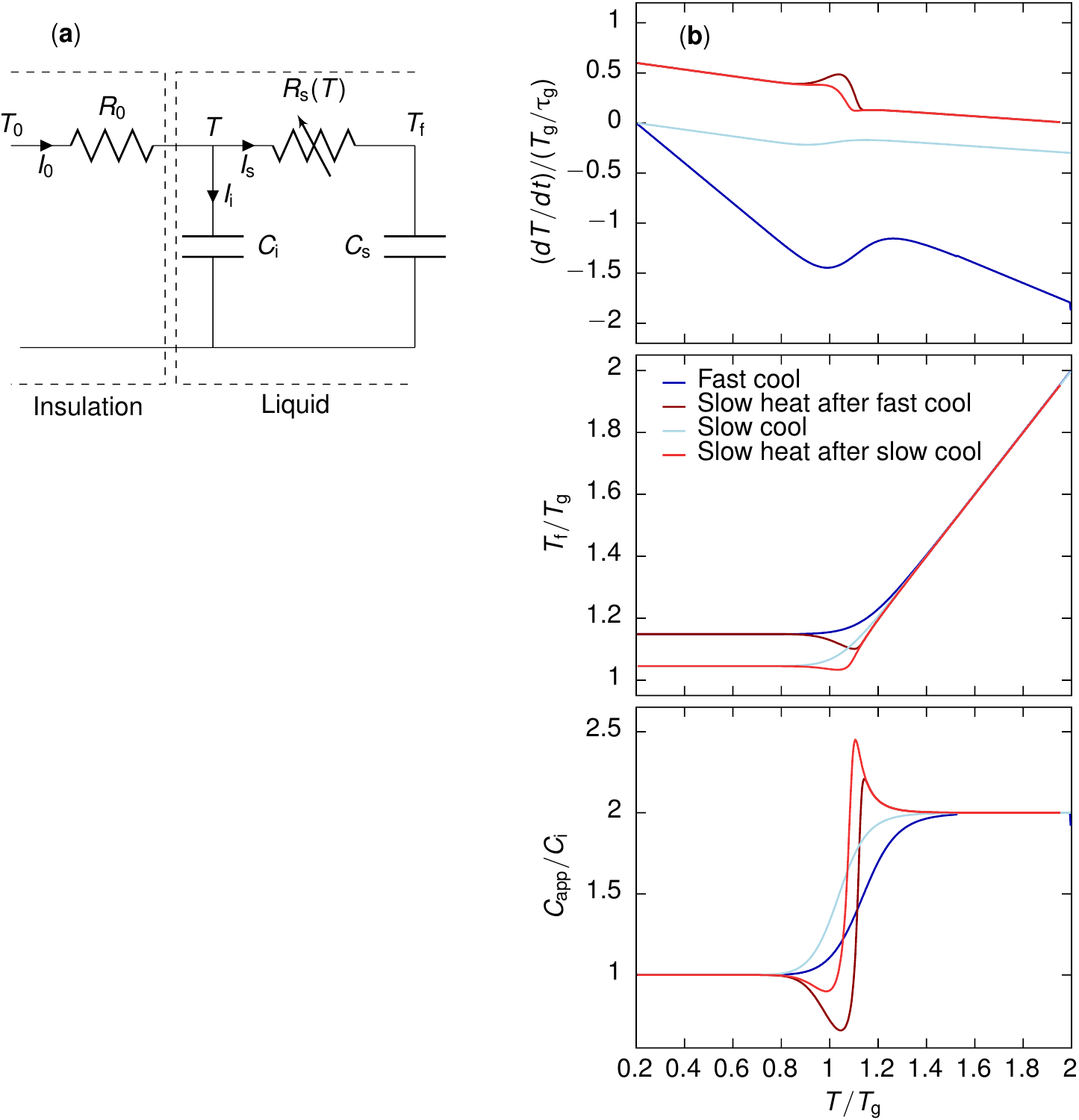}}
  \caption{The simple model for understanding the TC signal from a
    glass-forming liquid described in Appendix
    \ref{sec:simp-model-underst} (Eqs. (\ref{eq:7}) and
    (\ref{eq:model})). \textbf{(a)} Electrical network representation of
    the model. Voltages represent temperatures, electrical currents
    represent thermal currents, resistors represent thermal
    resistance, and capacitors represent specific heat. 
    \textbf{(b)} Results from model calculations by
    simple Euler integration using the implementation shown in Fig.\
    \ref{fig:modelcode} and the parameters given in Ref.\
    \onlinecite{Modparm}. $C_\mathrm{app}$ is the apparent specific heat as
    it would be observed by the TC technique.  }
  \label{fig:model}
\end{figure}

\begin{figure}
  \includegraphics{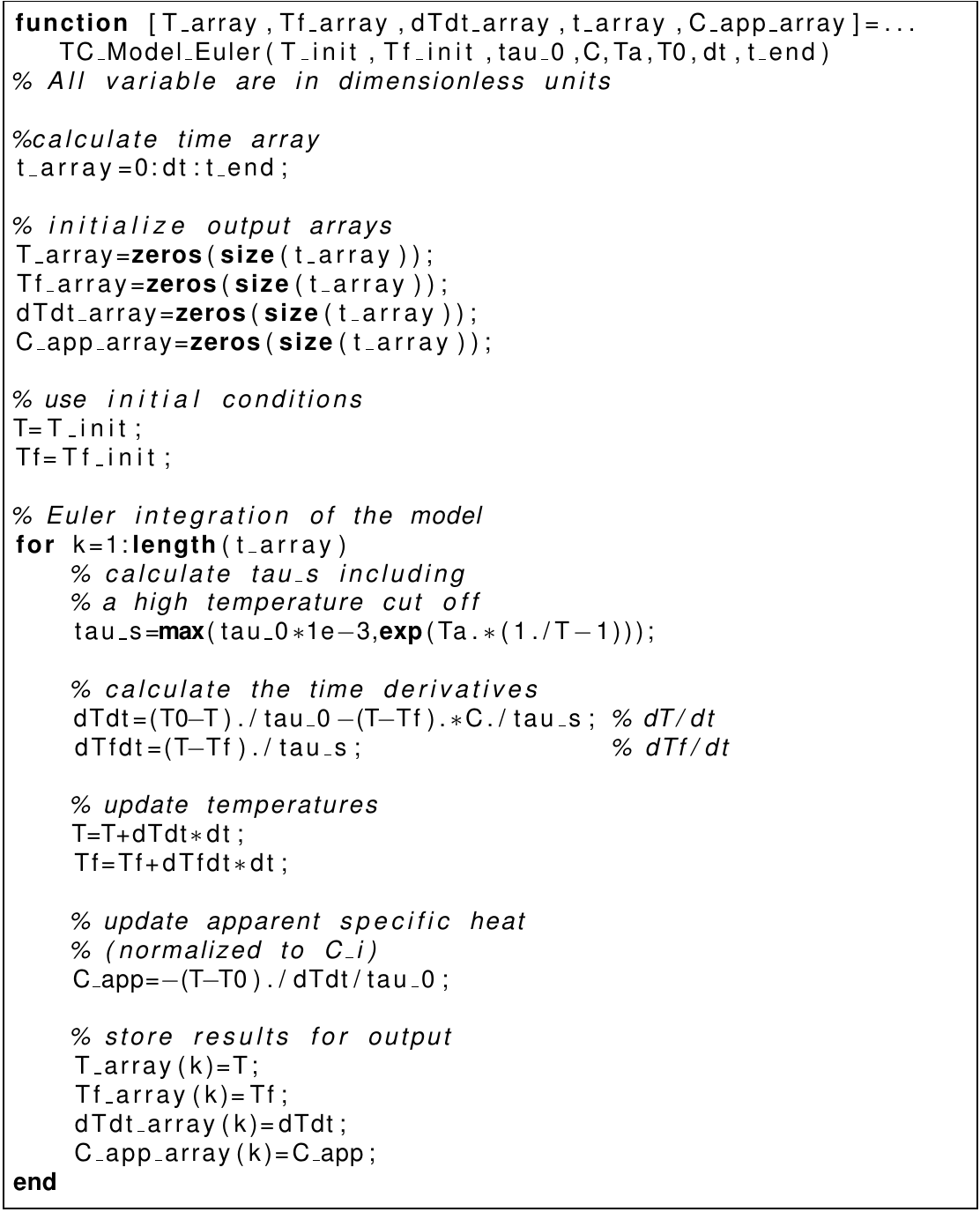}
  \caption{Matlab/octave code for numerical integration of the model
    given by Eq.\ (\ref{eq:simpmoddimless}) and illustrated in Fig.\ \ref{fig:model}(a).}
  \label{fig:modelcode}
\end{figure}

In this section we briefly discuss a simple model that can be used for
interpreting the results from the TC technique when applied to
glass-forming liquids. The model is based on the old understanding of
Tool and others \cite{Tool1946} that structure may be quantified
in terms of a so-called fictive temperature \cite{Dyre2006}. This is
the temperature at which the liquid's structure is identical to that
of the glass; thus for the metastable supercooled liquid the fictive
temperature, $T_\mathrm{f}$, is the actual temperature, whereas below
the glass transition $T_\mathrm{f}$ is constant, equal to the glass
transition temperature. For annealing just below the glass transition
the structure gradually changes and the fictive temperature moves slowly towards
the annealing temperature.

In the network representation of the model voltages correspond to
temperatures and electrical currents to heat currents. Specific heat
is represented as capacitors and thermal resistance as
resistors. Figure \ref{fig:model}(a) shows an electrical network
analog of the model and Fig.\ \ref{fig:model}(b) shows typical heating
and cooling curves obtained by solving the model.

The specific heat can be separated into two parts: an
``instantaneous'' contribution, $C_\text{i}$, which is able to take up
heat without structural rearrangement of the molecules; this
corresponds to the high-frequency limit of the dynamic specific heat,
and a structural contribution, $C_\text{s}$.

The major characteristic of a glass-forming liquid is that the time
scale separating the instantaneous and structural contributions to
relaxation phenomena increases dramatically when decreasing the
temperature \cite{Dyre2006}. This is reflected in the model by the
temperature-dependent resistor $R_\mathrm{s}(T)$, which in this simple
model is taken to be Arrhenius (generally, molecular glass-forming
liquids have non-Arrhenius  behavior \cite{Dyre2006,Hecksher2008}):

  \begin{align}
    \label{eq:7}
    R_\mathrm{s}(T)=R_\infty e^{T_\mathrm{a}/T}.
\end{align}

From the network it is seen that when the fictive temperature
($T_\mathrm{f}$) is identical to the actual temperature of the system
($T$) the system is in equilibrium and there is no current in the
$R_\mathrm{s}$ resistor, corresponding to no exchange
of energy between the instantaneous and the structural degrees of freedom.

In order to model the results of TC experiments the model also
includes the outer environment temperature, $T_0$, and the thermal
resistance to the surroundings, $R_0$, (according to Eq.\
(\ref{eq:power})).

The model can be expressed as a system of two non-linear Tool-type coupled
differential equations involving the sample temperature, $T$, and the
fictive temperature, $T_\mathrm{f}$:
\begin{subequations}\label{eq:model}
\begin{align}
    \diff{T(t)}{t}&=\frac{I_0-I_\mathrm{s}}{C_\mathrm{i}}=\frac{T_0-T}{R_0C_\mathrm{i}}-\frac{T-T_\mathrm{f}}{R_\mathrm{s}(T)C_\mathrm{i}}\\
    &= \frac{T_0-T}{\tau_0}-\frac{T-T_\mathrm{f}}{\tau_\mathrm{s}(T)}\tilde{C}\\
    \diff{T_\mathrm{f}(t)}{t}&=\frac{I_\mathrm{s}}{C_\mathrm{s}}=\frac{T-T_\mathrm{f}}{R_\mathrm{s}(T)C_\mathrm{s}}=\frac{T-T_\mathrm{f}}{\tau_\mathrm{s}(T)},
  \end{align}
\end{subequations}
where $\tau_0=R_0C_\mathrm{i}$ is the characteristic time of the
TC-technique, $\tau_\mathrm{s}(T)=R_\mathrm{s}(T)C_\mathrm{s}$ the 
characteristic relaxation time of the specific heat (separating
instantaneous and structural degrees of freedom), and
$\tilde{C}=C_\mathrm{s}/C_\mathrm{i}$ the ratio between structural and
instantaneous components to the specific heat. 

The structural characteristic time can be rewritten as:
\begin{align}
  \label{eq:9}
  \tau_\mathrm{s}(T)=C_\mathrm{s}R_\mathrm{s}(T)=C_\mathrm{s}R_\infty e^{T_\mathrm{a}/T}=\tau_\mathrm{g} e^{T_\mathrm{a}(1/T-1/T_\mathrm{g})}
\end{align}
where $T_\mathrm{g}$ is the glass-transition temperature and
$\tau_\mathrm{g}$ is the corresponding relaxation time. 

By choosing $\tau_\mathrm{g}$ and $T_\mathrm{g}$ as units for time and
temperature, respectively, the above set of equations can be brought to dimensionless
form:
\begin{subequations}\label{eq:simpmoddimless}
  \begin{align}
    \diff{\tilde T}{\tilde t}&=\frac{\tilde T_0- \tilde T}{\tilde
      \tau_0}-\frac{\tilde T-\tilde T_\mathrm{f}}{\tilde \tau_\mathrm{s}(T)}\tilde{C}\\
    \diff{\tilde T_\mathrm{f}}{\tilde t}&=\frac{\tilde T-\tilde T_\mathrm{f}}{\tilde
      \tau_\mathrm{s}(\tilde T)}\\
    \tilde\tau_\mathrm{s}(\tilde T)&=e^{\tilde T_\mathrm{a}\left(1/\tilde T - 1 \right)}
  \end{align}
\end{subequations}

The model described by Eq.\ (\ref{eq:simpmoddimless}) can be numerically
integrated by simple Euler integration. The Matlab function shown in
Fig.\ \ref{fig:modelcode} performs the integration and Fig.\
\ref{fig:model}(b) shows results from such model calculations.

This model is the simplest possible model for the specific heat of a
glass-forming liquid (it predicts a Debye-type equilibrium
frequency-dependent specific heat at variance with experiments (e.g.\
Refs.\ \onlinecite{Christensen1985,Birge1997}). The
model nevertheless captures qualitatively the shapes of the
\TdTdt{}-traces, both on cooling and heating, As well as the
difference between slow heating after slow and fast cooling, compare
Figs.\ \ref{fig:model}(b) and \ref{fig:measurement}(b).
 
%\vfill

%\pagebreak

%\bibliography{HRA}

\begin{thebibliography}{50}
\bibitem [{foo({\natexlab{a}})}]{footnote1}%
  \BibitemOpen
  \bibinfo {note} {For initial temperature of the sample $T_\text{initial}$ the
  solution of Eq.\ (\ref{eq:basic_eq}) is $T \left( t \right) ={T_0}+{{\rm
  e}^{-{\frac {t}{R_0C}}}} \left( {T_\text{initial}}-{T_0}
  \right)$}\BibitemShut {NoStop}%
\bibitem [{Sam()}]{Sampel}%
  \BibitemOpen
  \href@noop {} {}\bibinfo {note} {DC704 is the Dow Corning\textregistered{}
  diffusion pump oil: tetramethyl-tetraphenyl-trisiloxane. Glycerol
  (propane-1,2,3-triol) is a standard Flukar\textregistered{} glycerol.
  1-octyl-1-methylpyrrolidinium TFSI is the ionic liquid
  1-octyl-1-methylpyrrolidinium bis(trifuoromethanesulfonyl)imide acquired from
  Solvionic. Diethylene glycol is standard quality. Acetone is of standard
  laboratory solvent quality.}\BibitemShut {Stop}%
\bibitem [{foo({\natexlab{b}})}]{footnote5}%
  \BibitemOpen
  \bibinfo {note} {The timescales are defined as $\tau=1/(2\pi\nu_\text{lp})$,
  where $\nu_\text{lp}$ is the frequency at maximum loss in isothermal
  measurements of the frequency dependent complex response functions. Data on
  the characteristic timescale as function of temperature have been
  extrapolated down to $\tau=\SI{100}{\second}$ by linear extrapolation of
  $(T,\log{\tau})$ data.}\BibitemShut {Stop}%
\bibitem [{foo({\natexlab{c}})}]{footnote2}%
  \BibitemOpen
  \bibinfo {note} {Details, schematics and software associated with the
  TC-implementations are available on request.}\BibitemShut {Stop}%
\bibitem [{foo({\natexlab{e}})}]{footnote4}%
  \BibitemOpen
  \bibinfo {note} {The amplifier utilized for the digital TC implementation is
  based on the Analog Devices AD8571 op-amp. This is a single supply (\SI
  {5}{\volt }), ultra low noise, zero drift op-amp. The op-amp was powered from
  the \SI {5}{\volt } supply on LTC2983 demo board. A standard inverting
  amplifier configuration was used with a passive low-pass filter on the input
  and $255.5$ times amplification.}\BibitemShut {Stop}%
\bibitem [{Mod()}]{Modparm}%
  \BibitemOpen
  \href@noop {} {}\bibinfo {note} {Model parameters used for the integration of
  Eq.\ \ref{eq:simpmoddimless} to give the results presented in Fig.\
  \ref{fig:model}(b) are: $\tilde{T}_\text{room}=2$,
  $\tilde{T}_\text{cool}=0.2$, $\tilde{C}=1$, $\tilde{T_\text{a}}=20$,
  $\tilde{\tau}_{0,\text{fast}}=0.5$, $\tilde{\tau}_{0,\text{slow}}=3$. For
  cooling, the system is assumed to be initial in equilibrium
  ($\tilde{T}=\tilde{T_\text{f}}=\tilde{T}_\text{room}$). The final values from
  the cooling is used as initial condition for the heating. The integration was
  performed using the function presented in Fig.\ \ref{fig:modelcode} using a
  time step size of $0.0001$.}\BibitemShut {Stop}%
\end{thebibliography}


\end{document}